\documentclass[12pt]{article}

\input epsf

\def\d{\mbox{d}}

\title{Precision measurement of the top quark mass from
$M_{b\ell}$ distribution in $t \to b \, \ell\nu$ decays
\footnote{Talk given at 12th Lomonosov Conference on Elementary
Particle Physics, Moscow, Russia, August 25-31, 2005.}}
\author{M. L. Nekrasov \\
{\small\it Institute for High Energy Physics, 142284 Protvino,
Russia}}

\date{}

\begin{document}

\maketitle

\begin{abstract} The method of the top quark mass measurement from
$M_{b\ell}$ distribution in $t \to b \, \ell\nu$ decays can be
considerably improved if applying the technique of moments and
proceeding to moments of high degree. At the LHC this allows one
to reduce the systematic and statistical errors of the top mass by
more than a factor of two.
\end{abstract}

The precision measurement of the top quark mass is necessary for
testing of the SM and/or selecting the probable scenario of its
extension. The current accuracy of the top mass, attained at
Tevatron, is $\Delta M_t = 2.9$~GeV, and a further progress is
expected mainly at LHC and a future International linear collider.
At the LHC the accuracy of measurement of the top mass is
anticipated at the level of 1--$\,$2 GeV and in view of large
statistics the main difficulties are expected from systematic
errors. An analysis of \cite{top} shows that the most promising
method from the point of view of optimization of the errors is
based on an investigation of the invariant-mass distribution of
$J\!/\!\psi + \ell$ system produced in decays $t \to bW \to b
\ell\nu$ with $b$-jet formation and the $J\!/\!\psi$ being a
member of the $b$-jet. The evaluation made by Monte-Carlo
modelling gives approximately 0.7~GeV for systematic error of the
top mass measured by this method and of about 1~GeV for
statistical error for 4 years of LHC operation \cite{Kharchilava}.
In this paper we discuss an opportunity to improve this result by
means of optimization of the data handling within the same
experiment.

So, following \cite{Kharchilava} we consider the processes
\begin{equation}\label{Eq1}
q \bar q,\, gg  \to t \bar t \to b W \, b W \to b \ell \nu \,\, b
q_1 q_2 \Rightarrow \{ b\mbox{-jet} + \ell\} + \{3 \mbox{
jets}\}\,,
\end{equation}
with the $b$-jet and the isolated lepton $\ell = \{e,\mu\}$ coming
from one $t$ quark, and the remaining three jets coming from
another $t$ quark. In the experiment the above states are
registered and the invariant-mass distribution of $J\!/\!\psi +
\ell$ system is measured. The $M_{J\!/\!\psi \, \ell}$
distributions can then be compared with a template of shapes
parametrized by the top mass and thus $M_t$ can be fitted
\cite{Kharchilava}. One~could note, however, that the
invariant-mass distribution of the $b + \ell$ system inevitably
emerges at a certain stage of this analysis. Therefore on the
equal rights one can carry out the fitting in terms of the data
converted to the form of $M_{b\ell}$ distribution. In what follows
we adhere to the latter option and hence consider that (as if) a
distribution $F(q) = \sigma^{-1} \, \d \sigma / \d q$ is measured,
where $\sigma$ is the cross-section of the process (\ref{Eq1}) and
$q=M_{b\ell}$ is the reconstructed invariant mass of the $b +
\ell$ system.

By proceeding in this manner, we simulate the data under the
following sup\-positions. First we suppose that there is a
satisfactory method for extracting signal from the data. (Actually
this means the existence of a satisfactory~model for background
processes that survive after setting of kinematic cuts
\cite{Kharchilava}.) Secondly we describe the signal in the Born
approximation, identifying the $b$-jet with the $b$ quark.
Finally, on the basis of the results of \cite{Kharchilava}, we
disregard the effects of the finite width of the top quark. The
latter assumption means that $\sigma^{-1} \, \d\sigma/\d q$ is
equal to $\Gamma^{-1}_{b \ell \nu}\,\d \Gamma_{b \ell \nu} / \d
q$, where $\Gamma_{b \ell \nu}$ is a partial width of the decay $t
\to b\ell\nu$. (Thus the distribution $F$ becomes
process-independent.)

\begin{figure}
\hbox{ \hspace*{-5pt}
       \epsfxsize=380pt \epsfbox{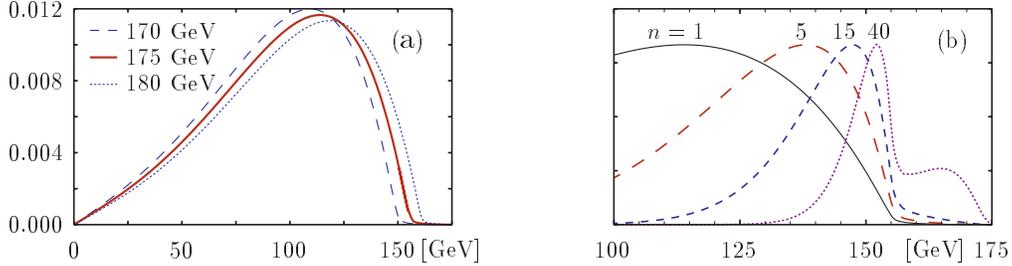}}
%
\caption{\footnotesize The distribution
$F(q)\!=\!\Gamma^{-1} \, \d \Gamma / \d q$, $q \equiv M_{b\ell}$,
at $M_t =$ 170, 175, 180 GeV (a), and the shape of $q^n F(q)$ at
$n=1$, 5, 15, 40 ($M_t \! = \! 175$ GeV), arbitrary normalization
(b).}\label{Fig1}
\end{figure}

At Fig.$\,$\ref{Fig1}(a) the distribution $F(q)$ is represented
for various values of the top mass. We see from the figure that in
the absolute value the distribution is most sensitive to $M_t$
in a region located between the maximum of the distribution and a
large-$q$ tail where $F(q)$ is almost vanishing. So in this region
of $q$-variable one could expect the highest accuracy for the
$M_t$ determination. A practical way to enhance the role of this
region is to proceed to high moments over the distribution,
\begin{equation}\label{Eq2}
\langle q^n \rangle = \int_0^{M} \!\!\! \d q \; q^n F(q)\,.
\end{equation}
Here $M$ is a technical parameter fixed close to $M_t$, and the
normalization of $F(q)$ is adjusted so as to satisfy the equality
$\langle 1 \rangle\!=\!1$. From Fig.$\,$\ref{Fig1}(b) it is
obvious that with increasing $n$ the distribution $q^n F(q)$ is
concentrating more and more in the above intermediate region,
increasing thus the sensitivity of $\langle q^n \rangle$ relative
to $M_t$.

So, by basing on the above observation, we define the
experimentally measured top-quark mass as a solution to the
equation
\begin{equation}\label{Eq3}
\langle q^n \rangle_{\mbox{\scriptsize exp}} = \langle q^n
\rangle\,.
\end{equation}
Here the moment in the l.h.s.~is determined on the basis of the
experimental data, and that in the r.h.s.~on the basis of the
theoretical distribution, which depends on the parameter $M_t$.
Let, for a given $n$, a solution to (\ref{Eq3}) be $M_t =
M_{t(n)}$. Then the error of the solution is
\begin{equation}\label{Eq4}
\Delta \! M_{t(n)} = \Delta \langle q^n \rangle_{\mbox{\scriptsize
exp}}\!\left/ \frac{\d \langle q^n \rangle}{\d
M_t}\Bigl|_{M_t=M_{t(n)}}\Bigr. \right..
\end{equation}
Our aim is to estimate $\Delta \! M_{t(n)}$ and find an optimal
value of $n$ which would minimize $\Delta \! M_{t(n)}$. Since in
view of (\ref{Eq2}) the derivative $\d\langle q^n \rangle/\d M_t$
is known, the problem is reduced to the estimation of the error
$\Delta \langle q^n \rangle_{\mbox{\scriptsize exp}}$ or its
components, the statistical and systematic experimental errors.

The statistical error can be estimated, in almost
model-independent way, by means of the formula \cite{N}
\begin{equation}\label{Eq5}
\Delta^{\mbox{\scriptsize stat}}\langle q^n
\rangle_{\mbox{\scriptsize exp}} = \sqrt{\frac{1}{N}\langle q^{2n}
\rangle}\,.
\end{equation}
Here $N$ is the representative sample of events, and $\langle
q^{2n} \rangle$ is a theoretically calculated moment. For the
consistent determination of the systematic error the application
of a proper MC event generator is needed. The analysis
\cite{Kharchilava} made with the aid of the PYTHIA and HERVIG
events generators showed that the major uncertainties in the
$M_{J\!/\!\psi \, \ell}$ distribution were caused by ($i$) the
uncertainties in the $b$-quark fragmentation and ($ii$) the
background processes. It is obvious that at solving the inverse
problem, the reconstructing the $M_{b\ell}$ distribution from the
$M_{J\!/\!\psi \, \ell}$ one, the errors should be of the same
origin.

With this in mind we consider first the error resulting from the
uncertainty in the $b$-quark fragmentation. For brevity we call
this the type I error. At the level of the $M_{b\ell}$
distribution it appears as an uncertainty $\Delta q$ in the $q$
variable. Let us suppose that $\Delta q$ is sufficiently small,
and let us neglect nonlinear effects. Then we have
\begin{equation}\label{Eq6}
\Delta^{\mbox{\scriptsize sys I}}\langle q^n
\rangle_{\mbox{\scriptsize exp}} = \int_0^{M} \!\!\! \d q \; \,
\left[q^n  F(q)\right]^{\prime}\, \Delta q \,.
\end{equation}
Here the prime means the derivative with respect to $q$. The
$\Delta q$ is determined by the relation $\Delta q = r q$ with $r$
is a coefficient, $r \simeq \frac{1}{2} \Delta E_b/E_b$, where
$E_b$ is the energy of the $b$ quark in the laboratory frame and
$\Delta E_b$ is its error. In fact the mentioned relation for
$\Delta q$ follows from the observation that $q^2$ actually is a
doubled scalar product of 4-momenta of the $b$ quark and of the
lepton $\ell$ whose momentum is considered precisely determined.

The systematic error arising after subtraction of the background
processes---we call it the type II error---appears in the absolute
value of the distribution function. Denote the corresponding error
by $\delta F$, then we have
\begin{equation}\label{Eq7}
\Delta^{\mbox{\scriptsize sys II}}\langle q^n
\rangle_{\mbox{\scriptsize exp}} = \int_0^{M} \!\!\! \d q \; \,
q^n \, \delta F(q) \,.
\end{equation}
The function $\delta F(q)$ should be vanishing at the boundaries
of the phase space. We also assume that $\delta F$ only once
changes the sign when $q$ runs the values. The simplest function
satisfying to these requirements is a polynomial, $\delta F = h \,
q(q\!-\!M/2)(q\!-\!M)$ with parameter $h$ describing the amplitude
of the error.

Now we turn to numerical results. We assign the following values
for the parameters with global meaning: $M_W=80.4 \mbox{ GeV}$,
$\Gamma_W=2.1 \mbox{ GeV}$, $M_t \! = \! M \! = \! 175 \mbox{
GeV}$. The parameters $N$, $r$, $h$ take the values depending on
the conditions of observation. Since in \cite{Kharchilava} the
$M_{J\!/\!\psi}$ distribution was determined at $N =4000$ (with
kinematic cuts and for 4 years of LHC operation), we set this
value for $N$, as well. The parameters $r$ and $h$ may be fixed by
basing on the properties of the $M_{J\!/\!\psi\, \, \ell}$
distribution. We omit here the respective discussion and show only
the result: $r \simeq 0.005$, $h \simeq 1.7 \times 10^{-10}$
GeV$^{-4}$ \cite{N}. Note that the same estimation for $r$ follows
from the above mentioned formula $r \simeq \frac{1}{2} \Delta
E_b/E_b$ when taking into consideration the 1\%-precision of the
determination of the energy of the $b$ jets expected at LHC
\cite{top}.

\begin{figure}
\hbox{ \hspace*{5pt}
       \epsfxsize=370pt \epsfbox{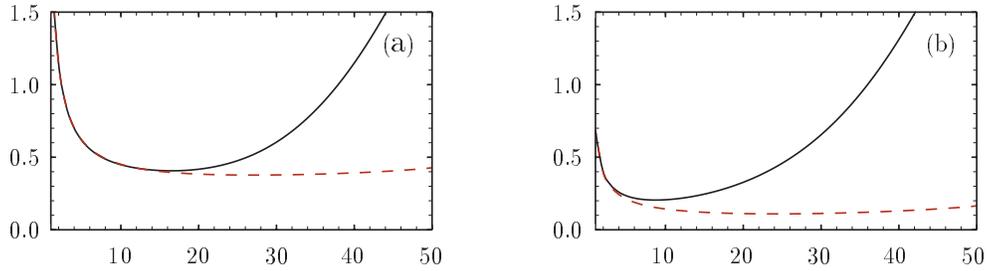}}
\vspace*{-3mm}\caption{\footnotesize The
$\Delta^{\mbox{\scriptsize stat}} M_{t(n)}$ (a) and
$\Delta^{\mbox{\scriptsize sys}} M_{t(n)}$ (b) in GeV depending on
$n$.}\label{Fig2}
\end{figure}

Now, as we know the parameters, we can calculate the errors of the
top~mass. At Fig.$\,$\ref{Fig2}(a) we show the behavior of
$\Delta^{\mbox{\scriptsize stat}} M_{t(n)}$ versus degree $n$ of
the moments. Fig.$\,$\ref{Fig2}(b) shows the behavior of the
summed-quadrature systematic~error $\Delta^{\mbox{\scriptsize
sys}} M_{t(n)}$. The dashed lines in both figures represent the
errors obtained~with usage of effective moments determined by way
of introducing a cut-off $\Lambda_n$ instead of $M$ in formulas
(\ref{Eq2}), (\ref{Eq6}) and (\ref{Eq7}); the aim of introducing
the cut-off is to isolate the tail of $q^n F(q)$ with an
unphysical peak arising at large $n$, see Fig.$\,$\ref{Fig1}(b).
Having the errors $\Delta^{\mbox{\scriptsize stat}} M_{t(n)}$ and
$\Delta^{\mbox{\scriptsize sys}} M_{t(n)}$, it is easily shown
that near $n=15$ the total error reaches the minimum $\Delta
M_{t(15)}=0.48 (0.42)$ GeV, with $\Delta^{\mbox{\scriptsize stat}}
M_{t(15)}=0.41 (0.40)$ GeV and $\Delta^{\mbox{\scriptsize sys}}
M_{t(15)}=0.24 (0.08)$ GeV. Here we put into the brackets the
errors calculated by the effective-moment method. It can be shown
by the naive counting method \cite{N} that the theoretical errors
will not spoil the above accuracy provided that the $M_{b\ell}$
distribution will be calculated within the one-loop electroweak
and two-loop QCD precision.

In summary, we discussed the potentialities inherent in the
technique of mo\-ments and demonstrated that at the LHC the
proceeding to moments of high degree allows one to reduce the
total error of $M_{t}$ to the level of about 500 MeV.

\end{document}